# Latent Networks Fusion based Model for Event Recommendation in Offline Ephemeral Social Networks

Guoqiong Liao, Yuchen Zhao, Sihong Xie, Philip S. Yu

## ABSTRACT

With the growing amount of mobile social media, offline ephemeral social networks (OffESNs) are receiving more and more attentions. Offline ephemeral social networks (OffESNs) are the networks created ad-hoc at a specific location for a specific purpose and lasting for short period of time, relying on mobile social media such as Radio Frequency Identification (RFID) and Bluetooth devices. The primary purpose of people in the OffESNs is to acquire and share information via attending prescheduled events. Event Recommendation over this kind of networks can facilitate attendees on selecting the prescheduled events and organizers on making resource planning. However, because of lack of users' preference and rating information, as well as explicit social relations, both rating based traditional recommendation methods and social-trust based recommendation methods can no longer work well to recommend events in the OffESNs. To address the challenges such as how to derive users' latent preferences and social relations and how to fuse the latent information in a unified model, we first construct two heterogeneous interaction social networks, an event participation network and a physical proximity network. Then, we use them to derive users' latent preferences and latent networks on social relations, including like-minded peers, co-attendees and friends. Finally, we propose an LNF (Latent Networks Fusion) model under a pairwise factor graph to infer event attendance probabilities for recommendation. Experiments on an RFID-based real conference dataset have demonstrated the effectiveness of the proposed model compared with typical solutions.

## Categories and Subject Descriptors

H2.8 [**Data Mining**]: Database Application – *data mining*

## 1. INTRODUCTION

With the growing amount of mobile social media, offline ephemeral social networks (OffESNs) are receiving more and more attentions. The OffESNs are defined as a kind of offline social networks created ad-hoc at a specific location for a specific purpose, e.g., hosting an academic conference or an expo, and lasting for some short period of time. The networks primarily rely on mobile social media such as Radio Frequency Identification (RFID) and Bluetooth devices to facilitate people connecting with each other freely[1-5].

Differing from online social networks, where people tend to share recent experiences and make friends with long-standing relationships, the OffESNs are face-to-face interaction and event-driven dynamic networks, where people are apt to share information and make temporary friends via physical proximity[6]. Some fundamental and important questions of the OffESNs still remain unexplored. For example, what are the factors to influence users' behaviors? Is there any latent social relation in the networks? Can we recommend events effectively by the latent social relations? And so on.

Basically, the primary purpose of people in the OffESNs is to acquire and share information via attending some prescheduled events such as keynotes, talks or exhibitions. It is an interesting and important work to make personalized recommendation over this kind of events. On the one hand, it can help attendees identify the potential conflict on multiple scheduling events and facilitate them making attendance decisions easily, since it is a common phenomenon in the networks that there are multiple events to be scheduled simultaneously due to the constraint of short period. On the other hand, it can also help event organizers guide resource planning. For example, if we can predict, based on the recommendation results, that the number of the people who will attend a future event is more than the capacity of its assigned room, the organizers can reassign a larger room for the event in advance.

*Motivating example.* Let's consider a scenario of an academic conference held in a specific hotel, where many people gather together for a few days, to attend keynotes or parallel sessions, or share personal experiences with their friends freely. This is a typically scenario of the OffESNs where people can make free face-to-face interactions frequently. For example, Bob is attending a conference, where there are 10 talks to be held during 10:00am~11:00am, on the first day. Our task is to recommend a ranked list of the 10 events to Bob by his individual features and the latent social relations. Moreover, we can provide the predicted numbers of people attending the future events to the organizers for resource allocation.

In the research domain of recommender systems, numerous studies have focused on item recommendation (e.g., movies, books, commodities). Such systems are widely implemented in e-commerce systems according to users' historical rating data[7]. Recently, with the popularity of social networks, social-trust based recommendation has recently been proposed to improve recommendation accuracy[8-11]. However, both of the two kinds of methods can no longer work well to recommend the events in the OffESNs due to following reasons.

- *Lack of users' preference and rating information.* Differing from the item recommendation in the Web, it is difficult to ask people to provide their real preference and give comments or ratings on the occurred events in the offline environment. Although we can retrieve a user's features such as research interests from the Web, we still don't know exactly which event is his/her favorite when he/she faces multiple events at the same time. Hence, the rating based traditional recommendation methods become invalid in the OffESNs.

---
*Guoqiong Liao is with the School of Information Technology, Jiangxi University of Finance and Economics, Nanchang 330013, China. E-mail: guoqiong.liao@gmail.com . This work was done when he visited UIC.*

*Yuchen Zhao is with Sumo Logic, Redwood 94063, USA. E-mail: zycemail @gmail.com. This work was done when he was a Ph. D. student at UIC.*

*Sihong Xie is with the Department of Computer Sciences, University of Illinois at Chicago(UIC), Chicago 60607, USA.E-mail: sxie6@uic.edu.*

*Philip S. Yu is with the Department of Computer Sciences, University of Illinois at Chicago(UIC), Chicago 60607, USA.E-mail: psyu@uic.edu.*

- *Lack of explicit social relations information.* In general, we usually don't know who old friends are or who will become new friends in the offline networks, so it is also difficult to obtain trusted social relations for recommendation. Zhuang et al.[2] used the online information such as co-authorships to facilitate predicting whether a pair of users will meet together in future time in a conference scenario. But the co-author relationships don't mean trusted relationships. Moreover, in some offline scenarios such as the expos, there are no co-authorships. Thus, the social-trust based methods cannot be used for event recommendation in the OffESNs effectively too.

Therefore, it is necessary to study new methods to recommend the events in the OffESNs. According to our observations, whether people attend the prescheduled events in the OffESNs are mainly determined by the factors in two aspects: individual interests or preferences on event contexts; social relations such as like-minded peers, co-attendees and friends, which are very useful for event recommendation. Unfortunately, we have little explicit information about them. Hence, there are two major challenges to recommend events in the networks:

- *How to derive users' latent preferences and the latent social relations from the observed historical interaction information in the networks?*

- *How to fuse multiple latent networks on social relations into a unified model to recommend the events effectively?*

To address the challenges, we first construct two observed heterogeneous interaction networks, an event participation network from human-event interactions and a physical proximity network from human-human interactions, to derive the latent preferences and the social relation networks. Then, we establish a fusion model to merge the latent networks to infer the probabilities that users will attend the future events. Finally, based on the probabilities, a ranked event list is recommended to each user.

To the best of our knowledge, this paper is the first work to define, construct and fuse multiple latent social networks for event recommendation in the offline social networks. Specifically, our contributions can be summarized as follows.

- We model the activities in the OffESNs as prescheduled events and spontaneous events, and construct two *observed heterogeneous interaction networks*: an event participation network and a physical proximity network.

- Based on the statistical measures on the two heterogeneous networks, we derive users' latent context preferences and three *latent social relation networks*, including a preference similarity network, an attendance relevancy network and an encounter network, to identify the social relations of like-minded peers, co-attendees and friends.

- We propose an effective Latent Networks Fusion (LNF) model based on a pairwise factor graph to fuse the three latent networks to infer the probabilities that users will attend the future events based upon their contexts, and provide a ranked event list for each user.

- Experiments on a real RFID-based conference dataset have demonstrated the effectiveness of our approach.

The rest of the paper is arranged as follows. We introduce the two heterogeneous interaction networks and problem statement in Section 2. In Section 3, we discuss our observations, and derive users' latent preferences and multiple latent social relation networks. We suggest a Latent Networks Fusion (LNF) model to fuse the three latent networks to infer event attendance probabilities and give an event ranking algorithm in section 4. In Section 5, we verify and evaluate the performance of the proposed model-based recommendation methods under a RFID-based real dataset. Section 6 is the related works, and we conclude the paper in the end.

## 2. HETEROGENOUS SOURCES AND PROBLEM STATEMENT

### 2.1 Heterogeneous interaction networks

In the OffESNs, people not only attend the prescheduled events, but also like to gather together spontaneously to talk about their common interests. Hence, the networks will generate a great deal of interaction information. In this work, we represent all activities in the networks as events.

**Definition 1** (*prescheduled events*). The prescheduled event is a scheduled social activity organized by the sponsors, e.g., an opening ceremony, a keynote or a talk. Such events usually have context information such as topics, time, locations, etc.

From the prescheduled events, we can obtain a lot of ephemeral human-event interaction information, e.g., event participation durations.

**Definition 2** (*spontaneous events*). The spontaneous event is a social activity initiated spontaneously by two or more people, such as having lunch together, talking about papers, news or else. Such events are unpredictable and are generated occasionally.

Differing from the prescheduled events, sometimes we don't know what people are actually doing in the spontaneous events. But from such events, we can obtain a great amount of sporadic human-human interactions (i.e., physical proximity) information, e.g., encounter durations and frequencies. Physical proximity is a significant metric to quantify users' offline behaviors[6]. For example, one may have much more physical proximity information with his friends than with other people.

Note that the recommendation task discussed in this paper is to recommend the prescheduled events. If there is no specific statement, the events mentioned in the paper indicate the prescheduled events.

**Definition 3** (*physical proximity*). If two or more people are involving in a spontaneous event for some period of time detected by the sensor devices, it is said that they are in physical proximity. i.e., they are encountering.

Physical proximity indicates the face-to-face interactions only generated from the spontaneous events, while the proximity due to attending the prescheduled events will be represented as another kinds of relations - co-attendees, which will be discussed in Section 3.1. In order to eliminate the noisy proximity information, it is required that the lasting time of each encounter should be longer than a duration threshold, which is determined by the specific scenario. In addition, although physical proximity is related to locations, here we put more attention on the facts that two or more people stay closely together. Therefore, the locations are not relevant information here.

Thus, from the above two kinds of events, we can establish two observed heterogeneous interaction networks: an event participation (human-event interaction) network and a physical proximity (human-human interaction) network.

Fig. 1(a) is an example of the event participation network to record the 4 users' participation history on the prescheduled events till $t$ is 2. In each time, 3 events were held simultaneously. The network is a bipartite graph including user nodes ($u_1 \sim u_4$) and event nodes ($e_{1.1} \sim e_{1.3}$, $e_{2.1} \sim e_{2.3}$). The edge attribute is the participation duration of the specific user on the specific event.

The example of the physical proximity network is shown as Fig. 1(b), which is used to record the encounter information between the users. The network only contains one kind of nodes – user nodes ($u_1 \sim u_4$). For a pair of users may encounter many times, it is possible for a pair to have multiple edges. The edge attribute in the network is the encounter duration.

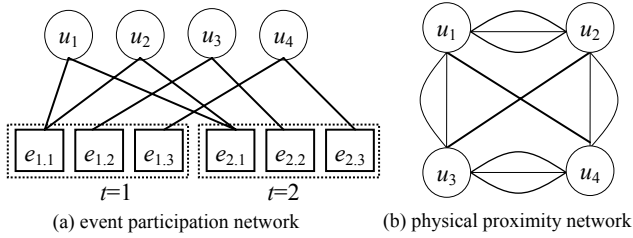

(a) event participation network   (b) physical proximity network

Figure 1. Observed heterogeneous interaction networks in the OffESNs

### 2.2 Problem statement

In this work, we use a part of data as the training set. Let $t_s$ be the begin time of the overall social event, $t_e$ be the end time, and $t \in [t_s, t_e]$. We consider the events during $[t_s, t]$ as training data and the events during $(t, t_e]$ as testing data for recommendation. We first define the input heterogeneous interaction networks formally, and then state our problem.

**Definition 4** (*event participation network*). The event participation network is denoted as $G_E^t = <U^t, E^t, X^t>$, where $U^t = \{u_i\}$ ($i=1, 2, .., N$) is the set of user nodes, $E^t = \{e_{j,k}\}$ ($j=1, 2, .., t$; $k=1, 2, .., M$) is the set of prescheduled event nodes occurring till time $t$, and an edge $x=(u_i, e_{j,k}) \in X^t$ represents the event participation relationship between $u_i$ and $e_{j,k}$, where the edge weight is the duration that $u_i$ attended $e_{j,k}$.

$G_E^t$ is a dynamic and incremental graph, where new edges will be added over time, i.e.

$$G_E^t = <U^t, E^t, X^t> = G_E^{t-1} \cup g_E^t = g_E^1 \cup g_E^2 \cup \ldots \cup g_E^{t-1} \cup g_E^t \quad (1)$$

where $g_E^t = <u^t, e^t, x^t>$ is an event participation sub-graph at time $t$. Of which, $u^t$, $e^t$ and $x^t$ are the sets of users, prescheduled events and participation relationships at $t$ respectively.

**Definition 5** (*physical proximity network*). The physical proximity network is denoted as $G_P^t = (U^t, Y^t)$, where $U^t$ is the set of user nodes and an edge $y=(u_i, u_j) \in Y^t$ represents the physical proximity relationship between $u_i$ and $u_j$, where the edge weight is the encounter duration.

**Problem 1.** Given $G_E^t$, $G_P^t$, and a set of prescheduled events $e^{t+1} = \{e_{t+1,1}, e_{t+1,2}, \ldots\}$ at time $t+1$. The goal of the work is to infer the probabilities that each user in $u_i \in U^t$ will attend each event $e_{t+1,j} \in e^{t+1}$, and rank the events based on the probabilities for personalized recommendation.

## 3. BUILDING LATENT NETWORKS ON SOCIAL RELATIONS

### 3.1 Observations

According to our observations, whether a user attends an event is influenced by following factors:

- His/her own preferences on the context of the event, i.e., explicit preferences
- The actions of the users having similar preferences, i.e., implicit preferences
- The actions of his/her friends

Thus, we can group the people in the OffESNs into three kinds of social relations as following:

- $SR_A$: The users who usually attend the events with the same contexts (they aren't required to attend the same events), called as *like-minded peers*.
- $SR_B$: The users who usually attend the same events together, called as *co-attendees*.
- $SR_C$: The users who usually encounter in the spontaneous events, called as *friends*.

However, both of the preferences and the social relations are latent in the OffESNs, we intend to use the observed heterogeneous interaction networks to derive them.

### 3.2 Interaction measures in OffESNs

An event refers to a real-world occurrence, which is described using the attributes such as who, where, when and what. The contextual attributes are application dependent[14]. For example, in the applications of location-based services (LBS), locations and time are critical aspects of their contexts. But in the OffESNs, for all prescheduled events are held in a building and last for a short period of time, people are more concerned with event topics than with the locations and time. Therefore, this work only considers the aspect of "what" as the context attributes. Note only the prescheduled events have the context attributes.

**Definition 6** (*event context*). We refer event topics as the contexts of the prescheduled events. The set of contexts is denoted as $C=\{c_i\}$, $i=1,2, \ldots d$, where $c_i$ represents the $i$-th context.

Thus we use the topics to represent the features of the prescheduled events. In general, an event can have more than one context, and a context can correspond to multiple events. Whether a user likes an event is determined whether he/she likes the contexts of the event. We first use the observed network $G_E^t$ to get the measures about context participation.

**Definition 7** (*context participation frequency*). The context participation frequency is the count that $u_i$ has attended the events with context $c_j$, which is equal to the number of the edges connecting $u_i$ and the events with $c_j$ in $G_E^t$, denoted as $PF_{ij}$.

**Definition 8** (*context participation time*). The context participation time is the sum of all durations that $u_i$ has attended the events with $c_j$, denoted as $PT_{ij}$, i.e.,

$$PT_{ij} = \sum_{k=1}^{PF_{ij}} PD_{ij}^k \quad (2)$$

where $PD_{ij}^k$ is the duration of $u_i$'s $k$-th attendance on $c_j$, which is equal to the weight of the corresponding edge in $G_E^t$.

We can also obtain the measures about encounter from the observed network $G_P^t$.

**Definition 9** (*encounter frequency*). The encounter frequency is the count that $u_i$ and $u_j$ have encountered in the spontaneous events, which is equal to the number of the edges between $u_i$ and $u_j$ in $G_P^t$, denoted as $EF_{ij}$.

**Definition 10** (*encounter time*). The encounter time is the sum of all durations that $u_i$ and $u_j$ have encountered in the spontaneous events, denoted as $ET_{ij}$, i.e.

$$ET_{ij} = \sum_{k=1}^{EF_{ij}} ED_{ij}^k \quad (3)$$

where $ED_{ij}^k$ is the duration of the $k$-th encounter, which is equal to the weight of the corresponding edge in $G_P^t$.

### 3.3 Derivation of latent information

Based on the measures mentioned above, this section first derives users' latent context preferences, and then derive and construct three latent networks on social relations including a preference similarity network, an attendance relevancy network and an encounter network, to identify the social relations such as like-minded peers, co-attendees and friends.

**Derivation of the latent preferences**

Because a user is usually interested in multiple contexts, we use a $d$-dimensional attribute vector $z_i$ to denote $u_i$'s latent preferences, where the $j$-th dimension attribute $z_{ij}$ represents the preference that $u_i$ on $c_j$.

Intuitively, the longer the time that a user attends an event is, the more interests he has on the event contexts. Thus, we intend to explore the context participation time to represent user's latent feedback.

**Definition 11** (*latent context preference*). The latent context preference $z_{ij}$ is the ratio between $u_i$'s participation time on $c_j$ and the total session time of $c_j$, i.e.,

$$z_{ij} = PT_{ij} / \sum_{c_j \in E^t} ST_j = \sum_{k=1}^{PF_{ij}} PD_{ij}^k / \sum_{c_j \in E^t} ST_j \quad (4)$$

where $ST_j$ is each session time of $c_j$.

**Derivation of the latent networks on social relations**

We can drive the latent social relation networks from the context preference, the observed networks $G_E^t$ and $G_P^t$.

For the overall context preferences of different users are difference, we use the adjusted cosine similarity[15] to represent users' preference similarity by subtracting the corresponding average context preferences.

**Definition 12** (*preference similarity*). The preference similarity is the measure to describe the similar degree based on context preferences between a pair of users $u_i$ and $u_j$, denoted as $\lambda_{ij}$:

$$\lambda_{ij} = \frac{\sum_{k=1}^{d}(z_{ik} - \overline{z_i})(z_{jk} - \overline{z_j})}{\sqrt{\sum_{k=1}^{d}(z_{ik} - \overline{z_i})^2} \sqrt{\sum_{k=1}^{n}(z_{jk} - \overline{z_j})^2}} \quad (5)$$

where $\overline{z_i}$ and $\overline{z_j}$ are the average preferences of $u_i$ and $u_j$, respectively.

For $\lambda_{ij} \in [-1, 1]$, we normalize it by Equation (6).

$$\lambda_{ij} = 1 - \frac{\cos^{-1}(\lambda_{ij})}{\pi} \quad (6)$$

**Definition 13** (*preference similarity network*). The preference similarity network is denoted as $G_S^t = (U^t, S^t)$, where $U^t$ is the set of user nodes, and each node $u_i \in U^t$ is associated with a $d$-dimensional attribute vector $z_i$, to denote $u_i$'s preferences. An edge $s = (u_i, u_j) \in S^t$ represents the preference similarity relationship, where the edge weight is equal to $\lambda_{ij}$.

**Definition 14** (*like-minded peers*). Let KNN($i$)\\$G_S^t$ be the $K$ nearest neighbors of $u_i$ in $G_S^t$. If $u_j \in$KNN($i$)\\$G_S^t$, it is said that $u_i$ and $u_j$ are like-minded peers.

In this work, we use the $K$-nearest-neighbor (KNN) method to determine the like-minded peers for the users. That is, for each user, we regard its $K$ highest preference similarity neighbors in $G_S^t$ as his/her like-minded peers. Therefore, we only need keep the connections between each node and its $K$ nearest neighbors in $G_S^t$.

In $G_E^t$, the more the neighbors (i.e., the common attendance events) that two users have, the more relevant they are. But the durations attending the same events of different users may be different, so we use a weighted Jaccard's Coefficient to calculate attendance relevancy.

**Definition 15** (*attendance relevancy*). The attendance relevancy is the measure to describe the correlations degree that a pair of users $u_i$ and $u_j$ attend events together, denoted as $\mu_{ij}$:

$$\mu_{ij} = \frac{\sum_{e \in \Gamma(u_i) \cap \Gamma(u_j)} D_i(e) + D_j(e)}{\sum_{e \in \Gamma(u_i)} D_i(e) + \sum_{e \in \Gamma(u_j)} D_j(e)} \quad (7)$$

where $D_i(e)$ and $D_j(e)$ are the durations that $u_i$ and $u_j$ attended the same event $e$ respectively, and $\Gamma(u_i)$ and $\Gamma(u_j)$ are the sets of events that $u_i$ and $u_j$ have attended respectively.

**Definition 16** (*attendance relevancy network*). The attendance relevancy network is denoted as $G_R^t = (U^t, R^t)$, where $U^t$ is the set of user nodes and an edge $r = (u_i, u_j) \in R^t$ represents the attendance correlation between $u_i$ and $u_j$, where the edge weight is equal to $\mu_{ij}$.

**Definition 17** (*co-attendees*). Let $\varphi$ be the threshold of the attendance relevancy. If $r = (u_i, u_j) \in R^t$ and $\mu_{ij} \geq \varphi$, it is said that $u_i$ and $u_j$ are co-attendees.

The observed network $G_P^t$ has recorded the encounter information between each pair, so we can be used to derive the encounter network.

**Definition 18** (*encounter network*). The encounter network is denoted as $G_Q^t = (U^t, Q^t)$, where $U^t$ is the set of user nodes and an edge $q = (u_i, u_j) \in Q^t$ represents the encounter relationship, where the edge weight is $EF_{ij}$ or $ET_{ij}$.

We can obtain two kinds of encounter networks by the weight used: frequency-based encounter networks using $EF_{ij}$ and time-based encounter networks using $ET_{ij}$. We will evaluate their influence on recommendation performance in Section 5.

By intuitions, the more count or time that a pair of users have encountered, the more closer they are, so that we can derive the friendship relations from the encounter network.

**Definition 19** (*friends*). Let $\delta$ be the threshold of the encounter frequency, $\theta$ be the threshold of the encounter time. If $q = (u_i, u_j) \in Q^t$ and $EF_{ij} \geq \delta$ in frequency-based encounter networks *or* $ED_{ij} \geq \theta$ in time-based encounter networks, it is said that they are friends.

# 4. EVENT RECOMMENDATION BASED LATENT NETWORKS FUSION MODEL

## 4.1 Basic idea

By now, we obtain three latent social relation networks, which have extracted all information from the observed heterogeneous interaction networks. Although now we can explore collaborative filtering (CF) methods[15-17] to recommend the events, since we can find the similar users from the preference similarity network. But in the OffESNs, besides users' individual preferences, users' actions are also influenced by the social relations, while the CF methods cannot handle such complexity.

The graphical models have long been used for modeling conditional dependency relationships between variables. An innovation work in [18] unified both directed and undirected graphical models as factor graphs, which provides a natural way of representing global functions or probability distributions that can be factored into simpler local functions, and is a widely used representation for modeling complex dependencies amongst hidden variables. Therefore, we intend to design a LNF model based on a pairwise factor graph (PGF) to infer the hidden probabilities that users will attend future events given the contexts. The symbols used in this section are listed in Table 1.

Table1. Notations

| Symbol | Description |
|---|---|
| $G_S^t = (U^t, S^t)$ | preference similarity network |
| $G_R^t = (U^t, R^t)$ | attendance relevancy network |
| $G_Q^t = (U^t, Q^t)$ | encounter network |
| $U^t=\{u_i\}(i=1, 2, .., N)$ | the set of users |
| $S^t$ | the set of edges in $G_S^t$ |
| $R^t$ | the set of edges in $G_R^t$ |
| $Q^t$ | the set of edges in $G_Q^t$ |
| $u_i$ | a user node or an observed variable |
| $z_i$ | $u_i$'s context preference vector |
| $z_{im}$ | the $m$-th attribute of $z_i$ |
| $Y=\{y_i\}(i=1, 2, .., N)$ | the set of hidden variables |
| $y_i$ | a single hidden variable |
| $\sim\{y_i\}$ | the set of variables in Y with $y_i$ removed |
| $y_{im}$ | the $m$-th attribute of $y_i$ |
| $g(y_i,u_i,KNN(i)\backslash G_S^t)$ | attributes feature function |
| $KNN(i)\backslash G_S^t$ | $u_i$'s K nearest neighbors in $G_S^t$ |
| $f(y_i, y_j)$ | attendance correlation function between $y_i$ and $y_j$ |
| $h(y_i, y_j)$ | local constraint function between $y_i$ and $y_j$ |
| $C=\{c_i\}(i=1, 2, .., d)$ | the set of event contexts |
| $c_m$ | the $m$-th context in C |
| $\lambda_{ij}$ | preference similarity between $u_i$ and $u_j$ |
| $\mu_{ij}$ | attendance relevancy between $u_i$ and $u_j$ |
| $EF_{ij}$ | encounter frequency between $u_i$ and $u_j$ |
| $ET_{ij}$ | encounter time between $u_i$ and $u_j$ |
| $\alpha_{ij}$ | indicator function of attendance correlations |
| $CM_{ij}$ | attendance correlation matrix between $y_i$ and $y_j$ |
| $CR_{ij}$ | behavior correlation between $u_i$ and $u_j$ |
| $\varphi$ | the threshold of attendance relevancy |
| $\delta$ | the threshold of encounter frequency |
| $\varphi$ | the threshold of encounter time |
| K | the number of the nearest neighbors |
| Z | a normalizing factor |

## 4.2 Latent networks fusion (LNF) model

The purpose of the latent networks fusion model is to incorporate the three latent networks on social relations into a unified model, to infer users' overall desires to attend a specific event, i.e., to predict how likely the users will attend a future event.

Figure 2 shows the fusion model under a pairwise factor graph implementation, where each factor function only involves two variable nodes. It is shown that all the three latent networks are captured via two types of variable nodes (the circle nodes in Fig. 2) and three types of factor function nodes (the square nodes in Fig. 2), which form the basic components of the model.

The corresponding relationships between the input latent networks and the PFG components can be seen clearly:

- The preference similarity network $G_S^t$ corresponds to observed variable nodes $U=\{u_i\}$ and attribute feature function nodes $g(y_i, u_i, KNN(i)\backslash G_S^t)$;
- The attendance relevancy network $G_R^t$ corresponds to the attendance correlation function nodes $f(y_i, y_j)$ and the related edges in the factor graph.;
- The encounter network $G_Q^t$ corresponds to the local constraint function nodes $h(y_i, y_j)$ and the related edges in the factor graph.

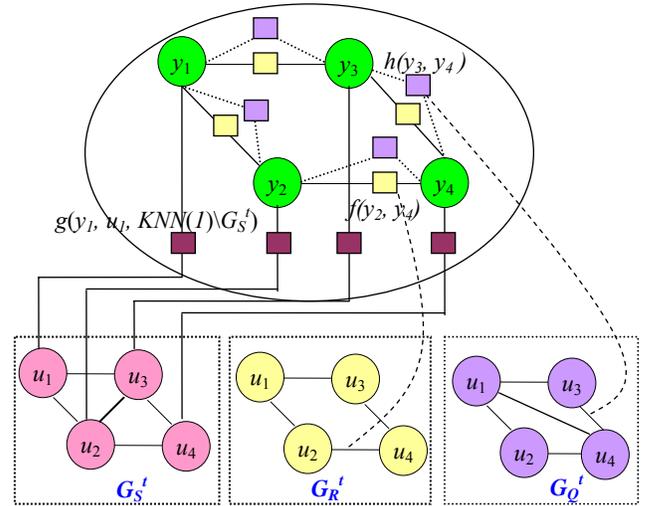

Figure 2. The pairwise factor graph implementation

## 4.3 Model definition

We give the definitions of the variable nodes and the factor function nodes, and design the concrete factor functions in this section.

**Definition 20** (*Observed variable nodes*). The observed variable nodes are the user nodes $U=\{u_i\}(i=1, 2, .., N)$ in $G_S^t$ (the pink circle nodes in Fig. 2), and $u_i$ has a $d$-dimension attribute vector $z_i$.

**Definition 21** (*Hidden variable nodes*). The hidden variable nodes $Y=\{y_i\}(i=1, 2, .., N)$ (the green circle nodes in Fig. 2) correspond to the $N$ observed variables nodes. Each $y_i$ is also a $d$-dimension attribute vector, and each dimension attribute is a binary discrete variables. For the $m$-th attribute $y_{im}$, the value as $y_{im}=1$ represents the probability that $u_i$ will attend context $c_m$; the value as $y_{im}=0$ represents the probability that $u_i$ will not attend context $c_m$.

**Definition 22** (*Attribute feature function nodes*). The attribute feature function nodes $g(y_i, u_i, KNN(i)\backslash G_S^t)$ (the brown square nodes in Fig. 2) define the correlations among the hidden variables $\{y_i\}$, $y_i$'s corresponding observed variables $u_i$ and $u_i$'s $K$ nearest neighbors in $G_S^t$, to represent the prior probabilities that $u_i$ will attend each context, e.g., $g(y_1, u_1, KNN(1)\backslash G_S^t)$ in Fig. 2.

**Definition 23** (*Attendance correlation function nodes*). The attendance correlation function nodes $f(y_i, y_j)$ (the yellow square nodes in Fig. 2) represent the attendance correlations between the hidden variables $y_i$ and $y_j$, e.g., $f(y_2, y_4)$ in Fig. 2.

**Definition 24** (*Local constraint function nodes*). The local constraint function nodes $h(y_i, y_j)$ (the pink square nodes in Fig. 2) reflect the constraint relations between the hidden variables $y_i$ and $y_j$, e.g., $h(y_3, y_4)$ in Fig. 2.

Basically, the attribute feature functions such as $g(y_i, u_i, KNN(i)\backslash G_S^t)$ describe the attributes of the hidden variable nodes, and the edge feature functions such as $f(y_i, y_j)$ and $h(y_i, y_j)$ describe the conditional dependency relations between the hidden variable nodes via edges.

*Attribute feature functions*

$g(y_i, u_i, KNN(i)\backslash G_S^t)$ is defined on the input network $G_S^t$. It incorporates $u_i$'s attribute vector $z_i$ and the attribute vectors of $u_i$'s $K$ nearest neighbors in $G_S^t$.

As the idea of the user-based collaborative filtering methods, the users with similar context preferences have similar ratings on the same things. We use the maximum value between $u_i$'s original preference $z_{im}$ and the weighted average preference of $u_i$'s $K$ nearest neighbors on context $c_m$ as the prior probability of $u_i$ attending $c_m$, denoted as $p_{im}$.

$$p_{im} = \max[z_{im}, \frac{1}{\sum_{j \in KNN(i)\backslash G_S^t} \lambda_{ij}} \sum_{j \in KNN(i)\backslash G_S^t} \lambda_{ij} z_{jm}] \quad (8)$$

The values of $y_i$'s $m$-th attribute can be computed as:

$$g(y_{im}, z_{im}, KNN(i)\backslash G_S^t) = \begin{cases} p_{im} & y_{im}=1 \\ 1-p_{im} & y_{im}=0 \end{cases} \quad (9)$$

*Attendance correlation functions*

$f(y_i, y_j)$ is an edge feature function to describe how likely that $u_i$ and $u_j$ attend the events together, i.e.,

$$f(y_i, y_j) = \alpha_{ij} * CM_{ij} \quad (10)$$

where $\alpha_{ij}$ is an indicator function to represent whether there is an edge between $u_i$ and $u_j$ in $G_R^t$ and the weight of the edge is more than $\varphi$, to filter the weak co-attendance relations.

$$\alpha_{ij} = \begin{cases} 1 & \mu_{ij} \geq \varphi, \text{ i.e., } u_i \text{ and } u_j \text{ are co-attendees} \\ 0 & \text{otherwise} \end{cases} \quad (11)$$

We represent the attendance correlations of each pair as a discretized correlations matrix $CM_{ij}$ as shown in Table 2, where $CR_{ij}$ represents the behavior correlation between $u_i$ and $u_j$ in different combinations, which is derived from the weight $\mu_{ij}$ in $G_R^t$.

Table 2. Definition of $CM_{ij}$

| $y_i$ | $y_j$ | $CR_{ij}$ |
|---|---|---|
| 0 | 0 | $\mu_{ij}$ |
| 0 | 1 | $1-\mu_{ij}$ |
| 1 | 0 | $1-\mu_{ij}$ |
| 1 | 1 | $\mu_{ij}$ |

The semantic of the matrix is: the cases of $y_i=0$, $y_j=0$ and $y_i=1$, $y_j=1$ indicate $u_i$ and $u_j$ take the same actions: for any event, either they don't attend, or they attend together; the cases of $y_i=1$, $y_j=0$ and $y_i=0$, $y_j=1$ indicate $u_i$ and $u_j$ take different actions: for any event, either $y_i$ attends, or $y_j$ attends.

*Local constraint functions*

We define the local constraint function $h(y_i, y_j)$ as an indicator function, taking friend relations in $G_Q^t$ as an input, i.e.,

$$h(y_i, y_j) = \begin{cases} 1 & EF_{ij} \geq \delta \text{ or } ED_{ij} \geq \theta, \text{ i.e., } u_i \text{ and } u_j \text{ are friends} \\ 0 & \text{otherwise} \end{cases} \quad (12)$$

Note the choice of $EF_{ij}$ or $ED_{ij}$ in the condition of Equation (12) is determined by the weight that the encounter network uses.

### 4.4 Probability inference and ranking algorithm

Based on the PFG implementation, we can formalize the context attendance probabilities as calculating the marginal probabilities of $y_i$ conditioned on the factor graph $G(Y)$ as:

$$p(y_i | G(Y)) = \sum_{\sim \{y_i\}} p(y_1, y_2, ... y_n | G(Y)) \quad (13)$$

where $p(y_1, y_2, ... y_n | G(Y))$ is the joint probability of all hidden variables in the PFG, which can be computed as Equation (13).

$$p(y_1, y_2, ... y_n | G(Y)) =$$
$$\frac{1}{Z}[(\prod_{i=1}^{N} \prod_{m=1}^{d} g(y_{im}, z_{im}, KNN(i) \backslash G_S^t)) \bullet (\prod_{i=1}^{N} \prod_{j=1}^{N} f(y_i, y_j) h(y_i, y_j))]$$
(14)

where $Z$ is a normalizing factor.

We use a widely used approximate iterative algorithm - loopy belief propagation (LBP)[19] to infer the joint probability distribution and the marginal probabilities. We omit the details due to the space limitation.

After obtaining the marginal probabilities, we can easily rank the parallel events as Algorithm 1.

| **Algorithm 1**. **Event-ranking ( )**. |
|---|
| INPUT: |
| $u^{t+1}=\{u_{t+1, i}\}$, $i=1,2, ..., n$; /* the set of users at time t+1*/ |
| $e^{t+1}=\{e_{t+1, j}\}$, $j=1, 2, ..., k$; /* the set of events at time t+1*/ |
| OUTPUT: |
| $\{\{e_i^{t+1}\}\}$, $i=1,2,..., n$; /* the set of ranked event lists */ |
| BEGIN |
| ① $C^{t+1}:=\varnothing$; /*initialization of the set of contexts*/ |
| ② for $j=1$ to $k$ |
| ③    $C_{t+1, j}:= GetContext(e_{t+1, j})$; /* get context of each event*/ |
| ④    $C^{t+1}:= C^{t+1} \cup C_{t+1, j}$ |
| ⑤ end for |
| ⑥ for $i=1$ to $n$ |
| ⑦    for $j=1$ to $k$ |
| ⑧       $p_{ij}=GetProb(p(y_{ij}=1))$ /*get marginal probabilities */ |
| ⑨    end for |
| ⑩    $\{e_i^{t+1}\}=RankDescend(p_{i1}, p_{i2}, ..., p_{ik})$ |
| ⑪    /* order by the probabilities in the descending order*/ |
| ⑫ end for |
| END |

# 5. EXPERIMENTS AND EVALUATIONS

## 5.1 Dataset description

The data set used for our experiments, "Attendee Meta-Data" (AMD), is downloaded from CRAWDAD (Community Resource for Archiving Wireless Data At Dartmouth)[20].

AMD HOPE was a project that aims to explore potential uses of RFID technology at "The Last HOPE" Conference held in July 18-20, 2008, New York, USA. It attempted to allow attendees to not only get a better conference experience, but also give them a new way to connect with other people.

All attendees in the conference received RFID badges that uniquely identified and tracked them across the conference area in the 3 days. RFID readers were deployed at 21 locations containing 3 conference rooms for tracking the attendees. The data set contained the information including attendee id numbers, interests, talks, tracking logs, etc. According to the talk schedule information, there were 39 parallel sessions and 99 talks. Each session contains 1~3 talks, which were held in the 3 conference rooms simultaneously lasting for 50 minutes. However, there are 5 sessions containing 15 talks without tracking logs, and 3 sessions had only 1 talk. We excluded these sessions. We thus left with 31 sessions containing 82 talks for the testing. In the experiments, we regard the talks as the prescheduled events, and incorporate the 21 interests provided by the dataset into 11 event contexts.

We divide the tracking logs into a training set and a testing set. The training set includes the tacking logs of the first 16 sessions containing 45 talks from which we obtain the event participation network, and other tracking logs in the same period from which we generate the physical proximity network. The testing set includes the latter 15 sessions containing 37 talks.

## 5.2 Data analysis and cleansing

For there are a lot of noisy data in the logs, we first clean them up based on statistical analysis results.

**Talk participation data analysis and cleansing**

There are 1,136,127 logs and 1,227 attendees in the initial tracking logs occurring in the 3 conference rooms in the 3 days. From the logs, we extract all talk participation records including an attendee id number, a talk id number and the attendance duration. It is found that the average participation duration is about 30 minutes.

The probability distribution of different duration intervals is shown in Fig. 3. We can see that about 20% of the durations are below 3 minutes. Since 3 minutes seems too short to reflect uses' real preferences, so we remove the participation records which durations are below 3 minutes.

Based on the above data, we analyze the probability distribution of the talk participation counts shown as in Fig. 4. There are over 25% attendees whose participation counts are less than 6. Because we use only half of the sessions for training, we remove the attendees whose participation counts in the training sessions are less than 3. Finally we are left with 915 attendees.

**Encounter data analysis and cleaning**

There are 21,374,278 tracking logs involving the 915 attendees occurring at the areas other than the conference rooms during the same period of the training sessions. From which, we extract all encounter records including for each pair (A, B), the ID numbers of A and B, and their encounter duration.

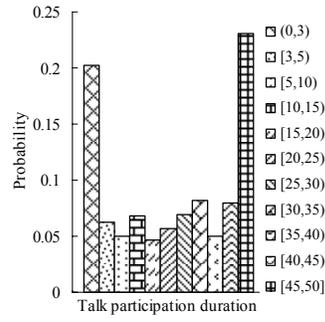 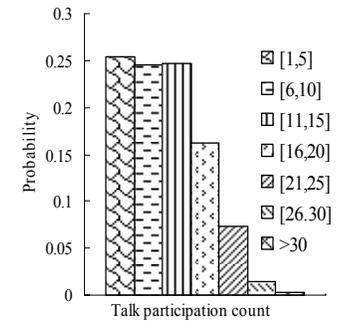

Fig.3  Probability analysis of talk participation durations

Fig.4 Probability analysis of talk participation counts

Fig. 5 demonstrates the probability distribution of the encounter durations. It shows almost 80% of the encounter durations are less than 3 minutes. After filtering them out, we obtain 36,573 encounter records.

We make further statistical analysis on the probability distribution of the encounter frequencies and the encounter time as shown in Fig. 6 and Fig. 7 respectively. The two figures show that about 80% of the encounter frequencies are only 1 and 80% of the encounter time are less than 10 minutes. Hence, the minimal threshold of the encounter frequency is set as 2, that of the encounter time set as 10 minutes. Because the cleansing process may also remove a lot of real encounter information, so the actual encounter frequencies and encounter time of each pair should be larger than the statistical results.

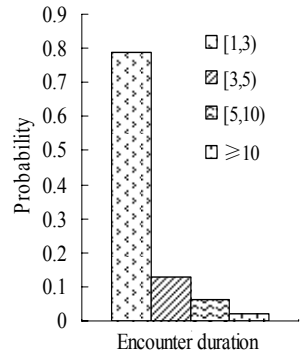 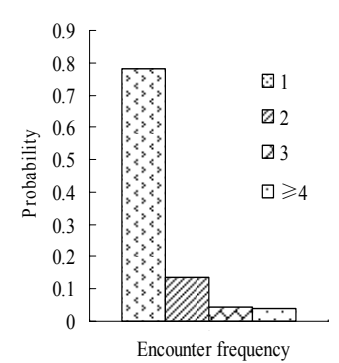

Fig.5  Probability analysis of encounter durations

Fig.6  Probability analysis of encounter frequencies

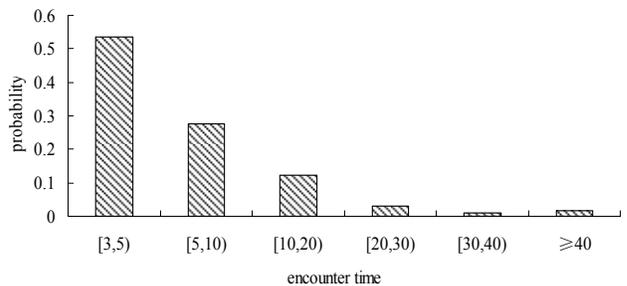

Fig.7  Probability analysis of encounter time

The experimental parameters are listed in Table 3.

Table 3. Experimental parameters

| Parameter | Description | Value |
|---|---|---|
| Cleansing condition | | |
| MIN(*PD*) | minimal talk presence duration | 3 min. |
| MIN(*PF*) | minimal talk presence frequency | 3 |
| MIN(*ED*) | minimal encounter duration | 3 min. |
| Network parameters | | |
| $N$ | number of user nodes | 915 |
| $M$ | number of event nodes | 45 |
| $|X|$ | number of edge in $G_E^t$ | 6,976 |
| $|Y|$ | number of edge in $G_P^t$ | 51,538 |
| $|R|$ | number of edge in $G_R^t$ | 313,138 |
| $|Q|$ | number of edge in $G_Q^t$ | 36,573 |
| Default thresholds | | |
| $K$ | threshold of number of neighbors | 6 |
| $\delta$ | threshold of encounter frequency | 6 |
| $\theta$ | threshold of encounter time | 30 min. |
| $\varphi$ | threshold of attendance relevancy | 0.4 |

## 5.3 Comparison methods and evaluation measures

The first goal of our evaluation is to examine whether the proposed model-based methods can improve recommendation performance. We also examine the contributions of different factors and the sensitivity of thresholds.

For we have three factors and two kinds of encounter networks, we will test following LNF models:

- **LNF-g**: The LNF model only using the *g* factor, i.e., the $G_S^t$ network.

- **LNF-gf**: The LNF model using both factors *g* and *f*, i.e., the $G_S^t$ and $G_R^t$ networks.

- **LNF-gfh-EF**. The LNF model using all three factors based on the frequency-based encounter network, i.e., all three latent networks.

- **LNF-gfh-ET.** The LNF model using all three factors based on the time-based encounter network, i.e., all three latent networks.

Two baseline methods are used for comparison:

- **Naïve method**. This method ranks the talks only by context preferences derived from the event participation network. Hence, this method only uses the node attributes in $G_S^t$ for recommendation.

- **User-based collaborative filtering** (**UBCF**) method[21]. Because of the widespread usage, we choose the user-based nearest neighbor algorithm as another baseline algorithm. For a user, if its preference value of a specific context is 0, the UBCF method is used. This method uses the node attributes and neighbor relations in $G_S^t$ for recommendation, and 5% nearest neighbors are determined.

In the test, we generate a ranking list of the parallel talks in each session for each user. The results are evaluated against users' real actions (i.e. ground truths) according to the metrics of precision and normalized discounted cumulative gain (nDCG)[22].

The precision is used to test the correct ratio of only recommending the talks with the maximum probabilities, which is the ratio between the number of correct recommendation and the number of all users' real participation actions.

A normalized discounted cumulative gain is a normalized version of a Discounted Cumulative Gain (DCG) measure that can account for differently output ranked lists, which is computed as[23]:

$$nDCG_p = \frac{DCG_p}{IDCG_p} \quad (14)$$

where IDCG is the ideal DCG, *p* is a particular rank position.

In general, a user attends one talk in a session, so the grades of all talks can be regarded as binary: 1 represents presence; 0 represents not-presence. Hence, we use Equation (15)[24] to compute the DCG, where $rel_i \in \{0, 1\}$, and $p \in \{2, 3\}$ since there are 2 or 3 talks in a session in the data set.

$$DCG_p = \sum_{i=1}^{p} \frac{2^{rel_i} - 1}{\log_2(i+1)} \quad (15)$$

For example, there are three talks *A*, *B* and *C* in a session. User $u_i$ attended *A*, but didn't attend *B* and *C*, so the grades of *A*, *B* and *C* for $u_i$ should be 1, 0 and 0, respectively. $IDCG_3$ is (1+0+0)=1. If we get a ranked list <*B*, *A*, *C*> for $u_i$, his $DCG_3$ is (0+0.63+0)=0.63, and $nDCG_3$ is 0.63/1=0.63.

## 5.4 Experiment results and discussion

**Performance comparison**

We first compare the recommendation performance between our LNF based methods and two baseline methods as shown in Figs. 8 and 9. It can be clearly seen that our methods outperform the baseline methods.

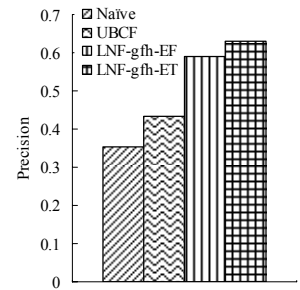 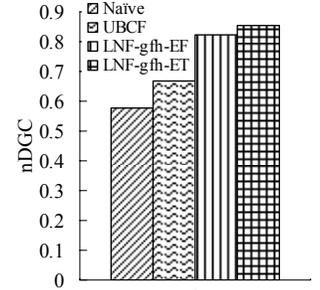

Fig.8 Precision comparison between our methods and the baselines

Fig.9 nDGC comparison between our methods and the baselines

In terms of the metric of precision, the LNF-gfh-EF and LNF-gfh-ET methods achieves improvements of approximately 16~20% compared to the UBCF method, and 24~28% to the naïve method. The two LNF methods also give a rise of 16~28% compared to the two baseline methods in nDGC. The reason of the improvements is that the LNF methods not only consider the preference similarity, but also take the attendance relevancy and friends relationships into account.

From the experimental results, we can find that LNF-gfh-ET can improve slightly the performance about 4% in precision and 3% in nDGC respectively compared to LNF-gfh-EF. It indicates the encounter time is more suitable to be used for representing the closeness between users than the encounter frequency.

**Factor contributions testing**

We also investigate the contributions of different factors as shown in Figs. 10 and 11.

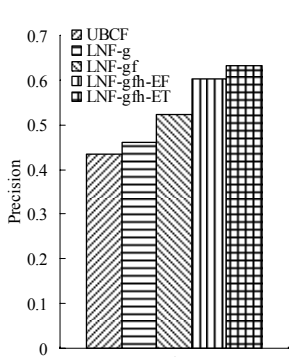
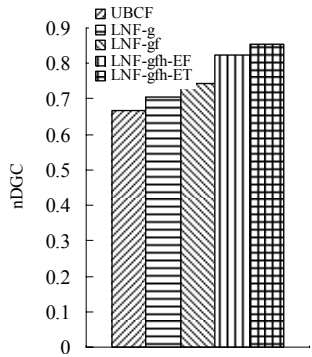

Fig.10 Precision comparison under different factors

Fig.11 nDGC comparison under different factors

We can see that the performances on both metrics of LNF-g (only uses factor *g*) are better than those of UBCF about 3%. This is because UBCF only considers neighbors' context preferences in the case that users' context preference values are 0, while LNF-g always considers neighbors' context preferences, sine it gets the maximum value between his own preferences and the weighed average preferences of the *K* nearest neighbors on the same context.

The LNF-gf method (uses factors *g* and *f*) can achieve performance improvements of approximately 4% compared to the LNF-g method. But its performances are worse than the LNF-gfh-EF and LNF-gfh-ET methods, since many users who often attend common talks may be irrelevant, which misleads the probability inference. This shows the importance of fusing all three latent networks.

**Thresholds sensitivity testing**

We further conduct experiments to investigate the sensitivity of different thresholds. Note when we test the performance of a specific threshold, the other thresholds are set the default values.

From Figs.12 and 13, we can see that the performance change trends both in precision and nDGC is stable when the *K* is up to 6.

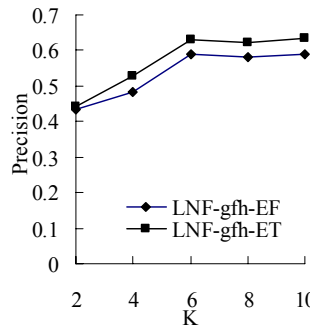
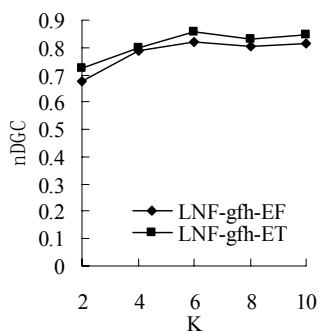

Fig.12 Precision comparison under different *K* values

Fig.13 nDGC comparison under different *K* values

Figs. 14 and 15 are the results when we select different thresholds of δ. We can see that 0.4 is the optimal point.

Fig. 16 is the test results on the LNF-gfh-EF method when we select different encounter frequencies (from 2 to 8). Similarly, Fig. 17 is the test results on the LNF-gfh-ET method when we select different encounter time (from 10 to 60 minutes). We can see that 6~8 is a better range for the encounter frequency, and 30~40 minutes for the encounter time.

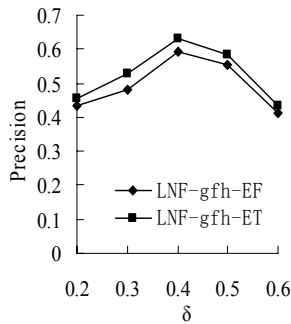
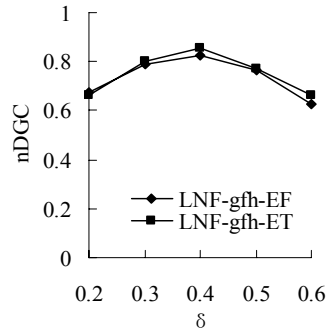

Fig.14 Precision comparison under different δ values

Fig.15 nDGC comparison under different δ values

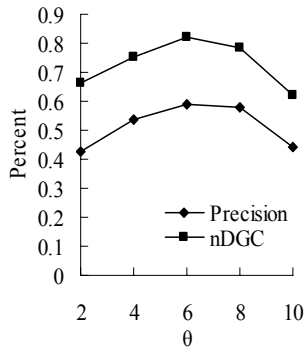
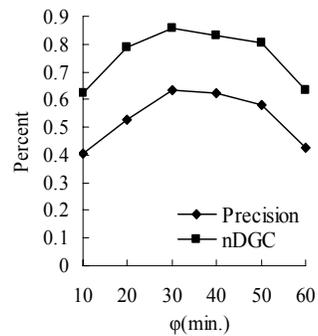

Fig.16 Precision and nDGC comparison under different θ values

Fig.17 Precision and nDGC comparison under different φ values

## 6. RELATED WORKS

**Traditional and social network recommendation**. Traditional recommendation systems mainly aim to recommend items that are likely to be interested to users. Such systems are widely implemented in e-commerce systems where the goal is to recommend items to the users by mining user rating history data. The methods used to solve this problem can be roughly categorized into content-based, collaborative filtering and hybrid approaches[7]. Recently, with the popularization of social networks, social-trust based recommendation has recently been proposed to improve recommendation accuracy[8-11]. The common rationale behind all of them is that a user's taste is influenced by her trusted friends in social networks[25]. Meanwhile, advances in location-based services and wireless communication technologies have enabled the creation of location-based online social networking such as Foursquare, Twinkle, and GeoLife[26-27], which make locations, users, activities and social media recommendation feasible[28-29]. However, because of lack of explicit users' preference, rating and trusted social relations in the OffESNs, both rating based traditional methods and social-trust based methods can no longer work well to recommend events in the OffESNs.

**Event recommendation.** The study on event recommendations is relatively little, especially in the offline social networks. For the Pittsburgh area, a cultural event recommender was build around trust relations[30]. Simon et al.[31] set up an online user-centric based evaluation experiment to find a recommendation algorithm that can improve user satisfaction for a popular Belgian cultural event website. Results show that a hybrid of a user-based collaborative filtering and content-based approach outperforms the other algorithms. Einat et al.[32] demonstrated a method for collaborative ranking of future events, which recommends the events based on individuals' preferences for past events, combined collaboratively with other peoples' likes and dislikes. But all these methods don't consider the characteristics of offline networks. Liu et al.[1] suggested event-based social networks (EBSN) containing both online and offline social interactions. It recommends events only based on the topological structures of the networks. The method can't be applied in the OffESNs since the networks usually can't get helps of the online interactions.

**Ephemeral social networks and prediction.** In recent two or three years, the offline ephemeral social networks have attracting the attentions of people. Alvin et al.[5] presented the concept of ephemeral social networks. It investigated how social connections can be established and examined user behaviors in the networks. Anne-Marie et al.[3] described the concept, potential applications, and underlying technologies of the offline social networks, and gave two scenarios of ephemeral social networks: remote meeting and family sharing. Christoph et al.[4] analyzed influence factors for link prediction and the strength of stronger ties in human contact networks, and used several network proximity measures to predict new links and recurring links. Zhuang et al.[2] formalized the problem of predicting geographic coincidences in the ephemeral social networks. It used a factor graph model integrating temporal correlations and social correlations to predict how likely two users will meet in future. However, both prediction methods are not event-driven and don't consider any context information when doing prediction, so they cannot be used for event recommendation in the OffESNs effectively.

## 7. CONCLUSIONS

In this work, we first construct two observed heterogeneous interaction networks including the event participation (human-event interaction) network and the physical proximity (human-human interaction) network. Based on them, we define some interaction measures in the OffESNs to construct the three latent social relation networks, including a preference similarity network, an attendance relevancy network and an encounter network. We then propose a latent networks fusion (LNF) model to merge the three latent networks into a pairwise factor graph to infer the probabilities that users will attend the future events given the contexts. The experimental results show that the suggested methods outperform the baseline methods.

An important contribution of our work is to propose a novel latent networks fusion based model for event recommendation in the offline ephemeral networks. The model uses a pairwise factor graph to factor the global functions into multiple simpler local functions, and capture all the features of the three latent networks via two types of variable nodes and three types of factor function nodes. The experimental results show the contributions of different factors, and verify the fusion model is effective.